\documentclass[a4paper,oneside,11pt,reqno]{amsart}
\usepackage{amssymb,bbm}
\usepackage{amsthm}
\usepackage{amsmath}
\usepackage{cite}
\usepackage[margin=2.7cm]{geometry}
\usepackage[foot]{amsaddr}

\usepackage{hyperref}
\hypersetup{
	pdftitle = {Dimension towers of SICs I: aligned SICs and embedded tight frames}, 
	pdfauthor = {Marcus Appleby, Ingemar Bengtsson, Irina Dumitru, Steven Flammia}
}

\expandafter\def\expandafter\normalsize\expandafter{%
    \normalsize
    \setlength\abovedisplayskip{12pt}
    \setlength\belowdisplayskip{12pt}
    \setlength\abovedisplayshortskip{10pt}
    \setlength\belowdisplayshortskip{10pt}
}

\makeatletter
\renewcommand{\@secnumfont}{\bfseries}
\renewcommand\section{\@startsection{section}{1}%
 \z@{.7\linespacing\@plus\linespacing}{.5\linespacing}%
  {\indent\normalfont\bfseries}}
\makeatother

\newtheorem{theorem}{Theorem}

\begin{document}

\vspace{1cm}

\begin{center}

{\Large DIMENSION TOWERS OF SICS. I.} 

~\\[4pt]

{\Large ALIGNED SICS AND EMBEDDED TIGHT FRAMES}

\vspace{1cm}

{\large Marcus Appleby}$^{*}$

\

{\large Ingemar Bengtsson}$^{\dagger}$

\

{\large Irina Dumitru}$^{\dagger}$

\

{\large Steven Flammia}$^{*,**}$

\

$^*${\small {\sl  Centre for Engineered Quantum Systems, School of Physics, \\
University of Sydney, Australia}}

\

$^{\dagger}${\small {\sl Stockholms Universitet, AlbaNova, Fysikum, \\
Stockholm, Sweden}}

\

$^{**}${\small {\sl  Center for Theoretical Physics, Massachusetts Institute of Technology, \\
Cambridge, USA}}

\vspace{7mm}

{\bf Abstract:}

\vspace{7mm} 

\parbox{117mm}{
\noindent Algebraic number theory relates SIC-POVMs in dimension $d>3$ to 
those in dimension $d(d-2)$. We define a SIC in dimension $d(d-2)$ to be 
aligned to a SIC in dimension $d$ if and only if the squares of the overlap 
phases in dimension $d$ appear as a subset of the overlap phases in dimension 
$d(d-2)$ in a specified way. We give 19 (mostly numerical) examples of aligned 
SICs. We conjecture that given any SIC in dimension $d$ there exists 
an aligned SIC in dimension $d(d-2)$. In all our examples the aligned SIC has 
lower dimensional equiangular tight frames embedded in it. If $d$ is odd 
so that a natural tensor product structure exists, we prove that the individual 
vectors in the aligned SIC have a very special entanglement 
structure, and the existence of the embedded tight frames follows as a theorem. 
If $d-2$ is an odd prime number we prove that a complete set of 
mutually unbiased bases can be obtained by reducing an aligned SIC to this 
dimension.} 
\end{center}

\newpage

\section{Introduction}\label{sec:intro}

\vspace{5mm}

\noindent It sometimes happens that an apparently simple question leads into 
very deep waters. We are concerned with just such a question here~\cite{Zauner, Renes}. 
To begin at the beginning, a SIC (also known as a 
SIC-POVM, or as a maximal complex equiangular tight frame) is a collection of $d^2$ unit 
vectors in ${\bf C}^d$ such that they resolve the identity, 
\begin{equation} \sum_{I=1}^{d^2} |\psi_I\rangle \langle \psi_I| = d 
{\mathbbm 1} \ , \label{resid} \end{equation}

\noindent and such that the absolute values of the overlaps $\langle \psi_I|\psi_J \rangle$ 
are equal (to $1/\sqrt{d+1}$ in fact) whenever $I \neq J$. The acronym stands for 
Symmetric Informationally Complete, and betrays the quantum state tomographical 
origin of the concept. In `Bloch space'---the affine space of Hermitian 
operators with unit trace equipped with the Hilbert-Schmidt inner product---a SIC 
is a maximal regular simplex, 
inscribed in the set of pure states. 
An obvious question is: Do SICs exist in all dimensions?
 
At the outset the SIC 
existence problem shows almost no structure. However, the known solutions make 
it clear that SICs are deeply implicated in a major open question in algebraic 
number theory. In every dimension that has been studied so 
far~\cite{Scott, Andrew, ACFW, FHS} there are SICs which are orbits under the 
discrete Weyl--Heisenberg group, a group with many applications in quantum 
mechanics~\cite{Weyl}, in radar and communication~\cite{HCW}, 
and in some approaches to Hilbert's 12th problem~\cite{Manin}. 
Remarkably, in every known example,
in the preferred basis singled out by the Weyl--Heisenberg group 
the components of the SIC vectors belong to abelian extensions of a real quadratic 
number field~\cite{AYAZ}. 
(We assume throughout that $d > 3$ and the SIC is Weyl--Heisenberg covariant.) 
Which real quadratic field that 
comes into play depends, contingent on a conjecture~\cite{AFMY}, in a known way on 
the dimension $d$. After a highly non-trivial but well understood extension of the 
quadratic field one arrives at a ray class field with conductor $d$ (or $2d$ if 
$d$ is even), and it appears that this always suffices to construct a SIC in dimension 
$d$~\cite{AFMY}. See ref.~\cite{AFMY2} for an account that assumes little or no 
background in number theory. Ray class fields are important because every abelian 
extension is contained in some ray class field. In many (presumably most) dimensions 
several unitarily inequivalent SICs exist, and further extensions of the ray class 
field are needed to construct them all. 

This particular connection between number theory and a simple geometric 
question was unexpected. It may be worthwhile to recall the connection between the 
geometry of regular polygons and the roots of unity. In number theoretic 
language the roots of unity generate extensions of the rational numbers, 
called cyclotomic fields. They are abelian extensions because the Galois group 
of the extension is abelian~\cite{Tignol}. Moreover the cyclotomic field 
generated by an $n$th root of unity is a ray class field over the rational 
number field ${\mathbb Q}$, with conductor $n$~\cite{Cohn}. 
The importance of the conductor is that one cyclotomic field is a subfield of 
another if the conductor of the one divides the conductor of the second. Every 
abelian extension of the rational numbers is a subfield of one of these ray class 
fields.  

A more pertinent example may be that of mutually unbiased bases (MUB) in dimensions 
$d$ such that $d$ is a prime number. Complete sets of such bases can be constructed 
using the Weyl--Heisenberg group, and in the preferred basis singled out by the 
group the components of all the MUB vectors can be constructed using $d$th 
roots of unity only (with a slight complication for $d = 2$)~\cite{Ivanovic}. 
Thus, to construct MUB in $d$ dimensions one needs cyclotomic fields with conductor 
$d$. Keep in mind that the roots of unity look extremely 
complex if one expresses them in terms of nested radicals, but they appear simple 
once it is realized that they can be obtained by evaluating the transcendental 
function $e^{2\pi i z}$ at rational points. (See Appendix~\ref{sec:roots}.) 
SICs are two orders of magnitude more difficult, 
because the relevant number fields are not yet fully understood. In particular, a 
description making use of special values of transcendental functions is conspicuously 
missing. Finding such a description forms an important part of the unsolved 12th 
problem on Hilbert's famous list. We say `two orders of magnitude' 
because there is a completed theory of abelian extensions of imaginary 
quadratic fields, one order of magnitude more difficult than the theory 
of the cyclotomic fields, and relying on the geometry of elliptic curves. Hilbert 
is reported as saying that this theory ``is not only the most beautiful part of 
mathematics but also of all science''~\cite{comedy}. But he wanted more, and 
understanding abelian extensions of the real quadratic fields seems a natural 
next step. 

We have reached the deep waters. To see how the dimension towers arise out of 
them, we need to add some details. The real quadratic field ${\mathbb Q}(\sqrt{D})$ 
conjectured to be relevant to SICs in dimension $d$ consists of the set of all 
numbers of the form $x + \sqrt{D}y$, where $x,y$ are rational numbers and~\cite{AYAZ}
\begin{equation} D = \mbox{square-free part of} \ (d+1)(d-3) \ . \end{equation}

\noindent Starting from this real quadratic number field one may perform further 
extensions to reach the ray class fields with conductor $d$ (or $2d$ if $d$ is even). 

The next question is what dimensions $d$ correspond to what square-free integers 
$D$. To see this one fixes a square free integer $D > 1$ and solves the 
Diophantine equation 
\begin{equation} (d+1)(d-3) = m^2D \hspace{5mm} \Leftrightarrow \hspace{5mm} 
(d-1)^2 - m^2D = 4 \end{equation}

\noindent for the integers $m$ and $d$. The solution consists of infinite sequences 
in each case~\cite{AFMY,AFMY2}. The beginnings of the sequences 
corresponding to the first three values of $D$ are  
\begin{align}
d&=7, 35, 199, 1155, 6727, 39203, 228487 \dots  & & \text{corresponding to} &D&=2
\label{seqD2}
\\
d&=5,15, 53, 195, 725, 2703, 10085 \dots & & \text{corresponding to} &D&=3
\label{seqD3}
\\
d&=4, 8, 19, 48, 124, 323, 844, \dots & & \text{corresponding to} &D&=5 \ . 
\label{seqD5}
\end{align}

\noindent The last of these sequences is noteworthy for the fact that it contains no less than 
seven dimensions less than $1000$, and is the subject of an important recent study by 
Grassl and Scott~\cite{GS}.    

\begin{figure}[t]
\begin{picture}(350,185)
\put(80,5){4}
\put(80,20){8}
\put(148,35){19}
\put(77,50){48}
\put(113,65){124}
\put(147,80){323}
\put(41,95){844}
\put(74,105){2208}
\put(3,117){5779}
\put(107,135){15128}
\put(83,15){\line(0,1){4}}
\put(83,30){\line(0,1){16}}
\put(83,60){\line(0,1){44}}
\put(83,117){\line(0,1){50}}
\put(153,90){\line(0,1){77}}
\put(153,45){\line(0,1){32}}
\put(119,75){\line(0,1){57}}
\put(119,146){\line(0,1){22}}
\put(48,105){\line(0,1){62}}
\put(13,126){\line(0,1){41}}
\put(88,14){\line(1,2){25}}
\put(87,28){\line(1,4){26}}
\put(79,15){\line(-1,3){26}}
\put(282,5){5}
\put(279,20){15}
\put(352,35){53}
\put(275,50){195}
\put(313,65){725}
\put(345,80){2703}
\put(241,95){10085}
\put(275,105){37635}
\put(204,117){140453}
\put(305,135){524175}
\put(285,15){\line(0,1){4}}
\put(285,30){\line(0,1){16}}
\put(285,60){\line(0,1){44}}
\put(285,117){\line(0,1){50}}
\put(358,46){\line(0,1){29}}
\put(358,91){\line(0,1){76}}
\put(323,76){\line(0,1){58}}
\put(323,145){\line(0,1){22}}
\put(253,105){\line(0,1){62}}
\put(218,126){\line(0,1){41}}
\put(291,14){\line(1,2){24}}
\put(290,29){\line(1,4){26}}
\put(280,15){\line(-1,3){26}}
\put(185,0){\line(0,1){174}}
\put(186,0){\line(0,1){174}}
\put(5,15){$D=5$}
\put(200,15){$D=3$}
\end{picture}
\caption{Ray class field inclusions for $D = 5$ and $D= 3$. A field at the 
upper end of a line contains the field at the lower end. When 
$d$ is even the conductor equals $2d$, but this does not affect the links. 
The intricate structure of the partially ordered set does not come through because only the 
ten lowest dimensions are shown. In this paper we will be concerned with the vertical 
connections only.}
\label{fig:ladders}
\end{figure}
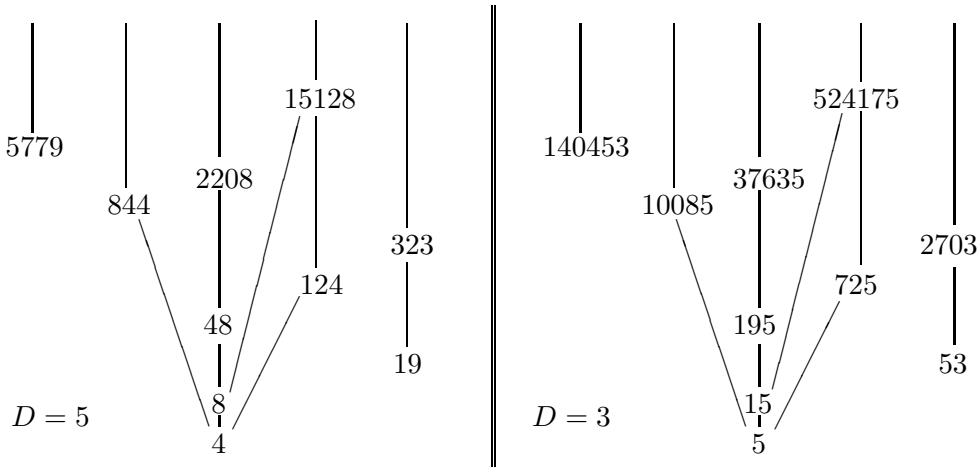

As with the cyclotomic fields, one field is a subfield of another if the conductor of 
the first divides the conductor of the other. Consequently, the divisibility properties 
of the dimensions give rise to an intricate partially ordered set ordered by field 
inclusions~\cite{AFMY, AFMY2}. See Figure \ref{fig:ladders}. Its structure is 
the same for each $D$. For instance, the first dimension in every sequence 
divides the second but not the third. In this paper we will be concerned with 
subsequences of the form $d_1, d_2 , \dots$ with the property $d_{j+1} = d_j(d_j-2)$ for 
all $j$. It is easily seen that the elements of such subsequences correspond to the 
same value of $D$. In fact, if $N = d(d-2)$ then 
\begin{equation} (N+1)(N-3) = (d^2 -2d+1)(d^2-2d - 3) = (d-1)^2(d+1)(d-3) \ . \end{equation} 

\noindent The square-free part is $(d+1)(d-3)$. 
Since $d$ divides $N$ the ray class field with conductor $d$ is a subfield 
of that with conductor $N$. The replacement $d \rightarrow d(d-2)$ thus generates an 
infinite `tower' (or `ladder') of 
ray class fields over the same real quadratic field, each one contained in the next.  
Examples of towers of this form include 
\begin{alignat}{9}
7& &\quad & \rightarrow & \quad & 35
& \quad & \rightarrow & \quad & 1155
& \quad & \rightarrow & \quad & \dots
& \qquad & \text{corresponding to}  & \qquad D&=2
\\
5& &\quad & \rightarrow & \quad & 15
& \quad & \rightarrow & \quad & 195
& \quad & \rightarrow & \quad & \dots
& \qquad & \text{corresponding to}  & \qquad D&=3
\\
4& &\quad & \rightarrow & \quad & 8 
& \quad & \rightarrow & \quad & 48
& \quad & \rightarrow & \quad & \dots
& \qquad & \text{corresponding to}  & \qquad D&=5 \ . 
\end{alignat}

\noindent As a glance at Figure \ref{fig:ladders} makes clear, there are other towers 
(such as $4 \rightarrow 124 \rightarrow 15128 \rightarrow \dots$) not 
considered here.

When translated into Hilbert space, this means that the number field from 
which one constructs $d$-dimensional SICs embeds into that used to construct 
$d(d-2)$-dimensional SICs. We are then led to ask how this number theoretic embedding 
manifests itself in terms of the geometry of Hilbert space. This question was first 
addressed by Gary McConnell, who studied the scalar products among SIC vectors and found  
that some of the overlap phases in dimension $d(d-2)$ actually belong to the smaller 
field. The pattern is subtle and has many facets. Here we focus on one of them: in 
every known example, we find that some of 
the overlap phases in dimension $d(d-2)$ are squares of overlap phases from dimension $d$, 
or the negative thereof. 
The precise relationship is described in Observations 1 and 2 in Section~\ref{sec:squared}.
This facet has significant geometrical consequences which we explore in Sections~\ref{sec:equi}--\ref{sec:sym}. 

This relationship between the phases leads to our definition of \emph{aligned} SICs, and we conjecture that corresponding to every SIC in dimension $d$ there is an aligned SIC in dimension $d(d-2)$. 
We observe that lower dimensional equiangular tight frames (ETFs) can be 
found embedded in all our examples of aligned SICs, as described in Section~\ref{sec:equi}. 

We then specialize to the case of odd dimensions. 
We study the entanglement properties of an aligned SIC in (odd) dimension $d(d-2)$, and prove two theorems regarding the spectrum of their reduced density operators in Section~\ref{sec:entangle}. 
We show that starting with an aligned SIC in dimension $p(p+2)$, for $p$ an odd prime, we can obtain a full set of MUB in dimension $p$ via an affine map; this is shown in Theorem~\ref{thm:3} in Section~\ref{sec:MUB}.  
We then show in Theorem~\ref{thm:4} in Section~\ref{sec:embed} that an aligned SIC in odd dimension $d(d-2)$ necessarily contains two ETFs of the kind whose existence was observed in Section~\ref{sec:equi}. 
Finally, we show in Theorem~\ref{thm:5} in Section~\ref{sec:sym} that such a SIC necessarily has the $F_b$ symmetry whose existence was noted empirically by Scott and Grassl~\cite{Scott, Andrew}.  

Proving the even dimensional analogs of the results proven in 
Sections~\ref{sec:entangle}--\ref{sec:sym} involves some significant complications, 
arising because in even dimensions $d$ and $d-2$ are not relatively prime. 
This case will be discussed in a subsequent publication. 

Our conclusions are given in Section~\ref{sec:conc}, where we also comment 
on the very recent and important results of Grassl and Scott~\cite{GS}.

\vspace{1cm}

\section{Preliminaries}\label{sec:prelim}

\vspace{5mm}

\noindent A Weyl--Heisenberg SIC in dimension $d$ is defined by a fiducial 
vector $|\psi_{0,0}\rangle$, from which the remaining SIC vectors 
$|\psi_{i,j}\rangle$ are obtained by acting with the $d^2$ displacement 
operators $D_{i,j}$. The labels are pairs of non-negative integers 
$0 \leq i,j < d$. For convenience these operators are often indexed by a 
two-component `vector' ${\bf p}$, and the SIC vectors are then written as 
$|\psi_{\bf p}\rangle = D_{\bf p}|\psi_{\bf 0}\rangle$. We use both 
notations interchangeably, guided by convenience rather than principle. 
Readers unfamiliar with these matters are referred to Appendix~\ref{sec:WH}, and readers 
who need to be convinced of the preferred role of the Weyl--Heisenberg group 
are referred to the literature~\cite{Zhu}. In dimension 8 there exists a 
sporadic SIC covariant under a related Heisenberg group. See ref.~\cite{Zhu2} for 
a recent discussion. It will be completely ignored here. 

The {\it SIC overlap phases} in dimension $d$ are defined by  
\begin{equation} e^{i\theta_{\bf p}} = \left\{ \begin{array}{ccl} 1 & \mbox{if} & 
{\bf p} = {\bf 0} \\ \\ \sqrt{d+1}\langle \psi_{\bf 0}|D_{\bf p}|\psi_{\bf 0}\rangle 
& \mbox{if} & {\bf p} \neq {\bf 0} \ . \end{array} \right. \label{theta} 
\end{equation} 

\noindent It turns out, in every case where an exact fiducial is known, that the overlap 
phases are algebraic integers, and in fact algebraic units, 
in the number fields they give rise to~\cite{AFMY,AFMY2}. In this respect they are similar 
to the roots of unity, which are algebraic units in the cyclotomic fields. 

The importance of the Weyl--Heisenberg group derives largely from the fact that it 
is a unitary operator basis~\cite{Schwinger}, which means that every operator $A$ acting 
on ${\mathbb C}^d$ admits a unique expansion 
\begin{equation} A = \sum_{\bf p} a_{\bf p}D_{\bf -p} \ , \hspace{8mm} 
a_{\bf p} = \frac{1}{d}\mbox{Tr}D_{{\bf p}}A \ . \label{expansion} \end{equation}

\noindent In particular, for a one-dimensional projector this specializes to  
\begin{equation} |\psi \rangle \langle \psi | = \frac{1}{d} 
\sum_{\bf p} D_{\bf -p} \langle \psi| 
D_{\bf p}|\psi \rangle \ . \end{equation}

\noindent This formula will enter most of our arguments. In particular it means that 
the vectors in a SIC can be reconstructed from their overlap phases. 

A technicality needs to be mentioned here, because it plays a large role in the 
intermediate stages of our argument. The choice of the fiducial vector---among the 
vectors in a given SIC---seems at 
first sight to be arbitrary, so that we might just as well consider the overlap phases 
\begin{equation} \langle \psi_{\bf q}|D_{\bf p}|\psi_{\bf q}\rangle = 
\langle \psi_{\bf 0}|D_{-{\bf q}}D_{\bf p}D_{\bf q}|\psi_{\bf 0}\rangle = 
\omega^{\langle {\bf p},{\bf q}\rangle }\langle \psi_{\bf 0}|D_{\bf p}|\psi_{\bf 0}\rangle \ , 
\end{equation}

\noindent where $\omega$ is a $d$th root of unity, $\langle {\bf p},{\bf q}\rangle $ is 
an integer modulo $d$, and we used properties of the displacement operators that 
are explained in Appendix~\ref{sec:WH}. But then the number theoretical properties 
of the overlap phases can get `polluted' by roots of unity. A good choice 
of the fiducial vector can be made by observing that the Clifford group (the 
unitary automorphism group of the Weyl--Heisenberg group) contains the symplectic 
group as a factor group. A definite copy of this group is represented by unitary 
operators $U_F$, where $F$ is a symplectic two-by-two matrix, with entries that 
are integers modulo $d$ (or $2d$ if $d$ is even)~\cite{Marcus}. It turns out, in 
every case where an exact or numerical fiducial is known, that there always exist special 
choices of $F$ and of the vectors such that $|\psi_{\boldsymbol{0}} \rangle$ is an 
eigenvector of $U_F$. Such SIC vectors are 
called {\it centred}. The SIC vector $|\psi_{\bf q}\rangle $ is left invariant by $D_{\bf q}
U_FD_{-{\bf q}}$, and is said to be displaced. Centred SIC vectors are our 
preferred fiducial vectors, because the overlaps then lie in a smaller field, and the 
action of the Galois group simplifies. In dimensions divisible by 3 there is a further 
complication, because then there are displacement operators commuting with the 
relevant $U_F$. As a result, centred SIC vectors come in triplets. It turns out, in 
every case where an exact fiducial is known, that one of them is singled out by the 
number theoretical properties of its overlap phases, and is said to be 
\emph{strongly centred}~\cite{AFMY,AFMY2}. 

We will need to distinguish SIC overlap phases in dimensions $d$ 
from those in dimension $N = d(d-2)$. The latter are defined, using a strongly 
centred SIC fiducial $|\Psi_{\bf 0}\rangle$ in dimension $N$, by 
\begin{equation} e^{i\Theta_{\bf p}} = \sqrt{N+1}\langle \Psi_{\bf 0}
|D_{\bf p}^{(N)}|\Psi_{\bf 0}\rangle 
= (d-1)\langle \Psi_{\bf 0}|D_{\bf p}^{(N)}|\Psi_{\bf 0}\rangle \ . 
\label{Theta} \end{equation} 

\noindent Again we set $e^{i\Theta_{0,0}} = 1$ by convention. We label the operators 
with a superscript to signify the dimension, whenever this is demanded for clarity. 
The other convention established here is that capital letters $\Theta$ and $\Psi$ are 
associated to the larger dimension $N$, whereas lower case $\theta$ and $\psi$ refer 
to overlap phases and fiducials in the smaller dimension $d$. 

Given that we know $ e^{i\theta_{\bf p}}$ in 
dimension $d$, what can we say about $e^{i\Theta_{\bf p}}$ in dimension $d(d-2)$? 
If there is a pattern, what are the geometrical consequences? We will present 
some theorems concerning the second question, but for a technical reason we will 
restrict ourselves to the case of odd dimensions $d$. The reason is that the 
integers $d$ and $d-2$ are relatively prime if the dimension $d$ is odd, 
and then the Weyl--Heisenberg group, and indeed the whole Clifford 
group, splits as a direct product. The Hilbert space ${\mathbb C}^{d(d-2)}$, with $d$ odd, 
is thus displayed as a tensor product ${\mathbb C}^d\otimes {\mathbb C}^{d-2}$ in a 
preferred way. The (known) details revolve around the Chinese remainder 
theorem from elementary number theory. They are spelled out in Appendix~\ref{sec:CRT}. The tensor 
product structure makes it much easier to describe the geometrical consequences that 
we have found. In particular we can then use the language of entanglement 
theory, and it is irresistible to make use of this when we can. We will prove that 
the entanglement properties of a SIC in $d(d-2)$ dimensions are very special if 
it is aligned to one in dimension $d$. Moreover, when 
$d-2$ is an odd prime number we can include mutually unbiased bases (MUB) in the 
picture, and we do so in Section~\ref{sec:MUB}.  

\vspace{1cm}

\section{Squared phases in dimensional towers}\label{sec:squared}

\vspace{5mm}

\noindent The observations that will lead to our definition of aligned SICs 
are summarized in Tables \ref{tab:fiducials3} and \ref{tab:fiducials}. 
Every SIC in the tables is aligned to the one immediately below it (if any), 
in a sense to be explained. Our calculations are numerical, and the precision 
limited. For $d \leq 15$ we used the numerical fiducials given 
by Scott and Grassl~\cite{Scott}.\footnote{In five cases exact calculations 
have been made by Gary McConnell.}  

Before presenting the tables, we make an important clarifying remark. It must be understood 
that {\it none} of the phenomena we describe in this section has been proved to be a necessary 
consequence of the definition of a SIC. Each property of SICs that we discuss in this section 
as being universal (i.e. holding for all SICs, assuming further yet unknown ones exist) should 
be read with the caveat, `in every known case'. Still, the claims are based on a large 
number of examples. At the end of this section we will frame a definition motivated by 
some of them. 

{\small 
\begin{table}[h]
\caption{{\small SIC ladders with three known rungs. Exactly known SICs 
are in boldface, and they are underlined if they are ray class SICs. The pair 
15ac are surrounded by brackets because they are constructed from the 
same field. The order of the symmetry group is given below the label, with 
an asterisk if anti-unitary symmetries are included, a subscript $a$ if the 
Zauner symmetry is of the unusual kind (see Eq. (\ref{Zaunermat}) for 
definitions), and a subscript $s$ if the fiducial sits in the smallest of the 
three Zauner subspaces, as explained further in the main text.\newline}}
\begin{center}
{\renewcommand{\arraystretch}{1.51}
\begin{tabular}
{||c|c||c|c|c|c||}
\hline
\underline{\bf 48g} & {\bf 48f} &  195d & 195b & 195a & 195c \\  
24$_a^*$ & 6 & 12 & 6 & 6 & 6 \\ \hline 
\underline{\bf 8b} & {\bf 8a} & \underline{\bf 15d} & {\bf 15b} & ({\bf 15a} & {\bf 15c}) \\
12$^*_s$ & 3 & 6 & 3 & 3 & 3  \\ \hline 
\underline{\bf 4a} &  & \underline{\bf 5a} & &  &  \\
6$^*$ &  & 3 & &  &  \\ \hline 
\hline 
\end{tabular}
}
\end{center}
\label{tab:fiducials3}
\end{table}
\begin{table}[h]
\caption{{\small SIC ladders with only two known rungs, with the same conventions 
as in the previous table.\newline
}}
\begin{center}
{\renewcommand{\arraystretch}{1.51}
\begin{tabular}
{||c||c|c||c|c||c||c|c|c||c|c||c|c||c||}
\hline  
	\underline{\bf 24c} & \underline{\bf 35j} 
	& {\bf 35i} & 63b & 63c & 80i & 99b & 99c & 99d & 120c & 120b & 143a & 143b & 168a \\
 6 & 12$^*_s$ & 6$_s$ & 
 6 & 6 & $6_s$ & 6 & 6 & 6 & $12_a$ & 6 & $6_s$ & $6_s$ & 6 \\ \hline 
 \underline{\bf 6a} & \underline{\bf 7b} & {\bf 7a}
& (\underline{\bf 9a} & \underline{\bf 9b}) & \underline{\bf 10a} & \underline{\bf 11c} 
& ({\bf 11a} & {\bf 11b}) & \underline{\bf 12b} & {\bf 12a} & (\underline{\bf 13a} & 
\underline{\bf 13b}) &
\underline{\bf 14b} \\
 3 & 6$^*$ & 3 & 3 & 3 & 3 & 3 & 3 & 3 & 6$_a^*$ & 3 & 3 & 3 & 3 \\ \hline 
\hline 
\end{tabular}
}
\end{center}
\label{tab:fiducials}
\end{table} 
}

First we should explain the labeling system used for SICs~\cite{Scott}. SICs in a 
given dimension fall into orbits of the extended Clifford group (see Appendix~\ref{sec:roots}), 
which includes both unitary and anti-unitary transformations. The number 
of such orbits varies with the dimension, in ways that are not yet understood. 
Every SIC is labeled by the dimension and a letter labeling the extended Clifford 
orbit to which it belongs. 

Every SIC vector is left invariant by a subgroup of the extended Clifford group, 
that also transforms the SIC into itself. For centred fiducials this symmetry 
group is a subgroup of the extended symplectic group. As suggested by a conjecture of 
Zauner's~\cite{Zauner}, and confirmed in all the examples, the symmetry 
group always contains a cyclic subgroup of order 3. It is generated by a unitary 
operator called the Zauner operator. 

For $d \leq 50$ the order of the symmetry group may increase with 
the labeling letter's position in the alphabet~\cite{Scott}. For higher dimensions no such 
system has been adopted. Then the lexicographical order reflects the order in which the 
various orbits were found~\cite{Andrew}. Thus 4a is on a unique orbit in dimension 4, 
48g has the highest symmetry of all SICs in dimension 48, and 63p is the last orbit 
that was discovered in dimension 63. If the labeling system reminds the reader 
of the labeling system used for spectral classes of stars (in logical order, 
OBAFGKM), then so be it. 

A striking fact is that the order of the 
symmetry group doubles for each rung of the ladder in the tables. 
The tables contain some extra information that can be ignored for the time 
being: In dimensions $d = 3$ or $6$ modulo 9 the symplectic group contains two 
different conjugacy classes of order 3 elements, represented by the matrices 
$F_z$ and $F_a$. See Eq.~(\ref{Zaunermat}). SICs invariant under $U_{F_z}$ exist 
in all dimensions, but if $d = 3$ modulo 9 SICs invariant under $U_{F_a}$ exist too. 
Being of order 3, the Zauner operators split the Hilbert space 
into three Zauner subspaces. SIC vectors are always to be found in the largest of 
these, but in dimensions $d = 8$ modulo 9 the smallest subspace also contains 
SIC fiducials. There holds
\begin{equation} \hspace{7mm} d = 3 \ {\rm or} \ 8 \ {\rm mod} \ 9 \hspace{5mm} 
\Leftrightarrow \hspace{5mm} d(d-2) = 3 \ {\rm mod} \ 9 \  \end{equation}
\begin{equation} d = 1 \ {\rm or} \ 4 \ {\rm or} \ 7 \ {\rm mod} \ 9 \hspace{5mm} 
\Leftrightarrow \hspace{5mm} d(d-2) = 8 \ {\rm mod} \ 9 \ . \end{equation}

\noindent Thus the first exceptional property is `inherited' by the next rung, 
the second is not. 

\begin{figure}[ht]
\begin{picture}(300,140)
\put(30,16){${\mathbb E}_{7b}$}
\put(34,27){\line(0,1){22}}
\put(30,50){${\mathbb E}_{7a}$}
\put(38,27){\line(2,3){26}}
\put(63,68){${\mathbb E}_{35j}$}
\put(38,61){\line(2,3){26}}
\put(67,78){\line(0,1){22}}
\put(63,104){${\mathbb E}_{35i}$}
\put(133,05){${\mathbb E}_{4a}$}
\put(142,14){\line(1,1){11}}
\put(155,26){${\mathbb E}_{8b}$}
\put(159,37){\line(0,1){22}}
\put(155,60){${\mathbb E}_{8a}$}
\put(161,37){\line(2,3){28}}
\put(186,81){${\mathbb E}_{48g}$}
\put(160,70){\line(2,1){24}}
\put(190,91){\line(0,1){22}}
\put(186,117){${\mathbb E}_{48f}$}
\put(246,09){${\mathbb E}_{5a}$}
\put(255,19){\line(1,1){11}}
\put(266,30){${\mathbb E}_{15d}$}
\put(270,41){\line(0,1){22}}
\put(266,64){${\mathbb E}_{15b}$}
\put(270,75){\line(0,1){22}}
\put(266,98){${\mathbb E}_{15ac}$}
\end{picture}
\caption{{\small Field inclusions in three of the towers. A field at an upper end of a line 
contains the field at the lower end. We walk up the ladders by stepping rightwards.}}
\label{fig:inclusions}
\end{figure}
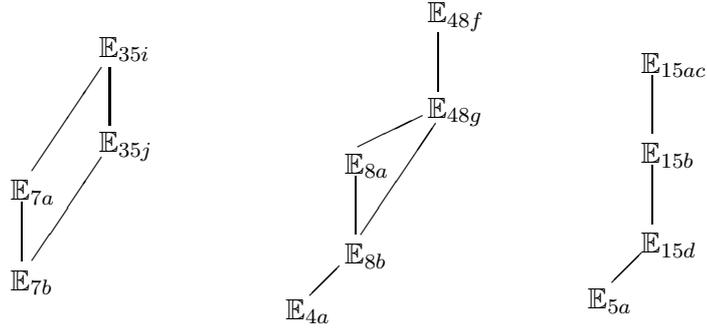

Each dimension contains a SIC known as a ray class SIC, constructed 
using a ray class field over the real quadratic field ${\mathbb Q}(\sqrt{D})$, 
where $D$ is the square free part of the integer $(d+1)(d-3)$.  
Other SICs in the same dimension are constructed from extensions of the ray 
class field. More precisely, there is a unique Galois multiplet (i.e. an orbit under 
the joint action of the Galois group and the extended Clifford group) of SICs belonging 
to the same ray class field; examples where the multiplet has more than one member 
include 9ab and 13ab~\cite{ACFW}. Field inclusions give rise to 
a partial ordering among the fields, 
given in Figure \ref{fig:inclusions} in the two cases where we have 
exact solutions available for more than one aligned SIC in the higher dimension. 
This pattern is not clear to us.

Our special concern in this paper is the phenomenology of squared SIC overlap phases. 
This can be summarized in two observations, relating some of the overlap phases in dimension 
$N = d(d-2)$ to those in dimension $d$:

\

\noindent {\bf First observation}. {\sl For SICs in dimension $d$ there exists 
a SIC in dimension $N = d(d-2)$, and a choice of fiducials, such that for ${\bf p} = (di,dj)$ 
we have} 
\begin{equation} e^{i\Theta_{di,dj}} = \left\{ \begin{array}{cll} 
+1 & \mbox{if} & d \ \mbox{is odd} \\ \\ -(-1)^{(i+1)(j+1)} & \mbox{if} & d 
\ \mbox{is even.} \end{array} \right. \end{equation} 

\

\noindent {\bf Second observation}. {\sl For SICs in odd dimensions $d$ there 
exists a SIC in dimension $N = d(d-2)$, and a choice of fiducials, such 
that $e^{i\Theta_{(d-2)i,(d-2)j}}$ is 
the negative of a square of an overlap phase from dimension $d$ if $d$ is odd. The 
relation between the phases is given by} 
\begin{equation} e^{i\Theta_{(d-2)i,(d-2)j}} = \left\{ \begin{array}{cll} 
-e^{2i\theta_{\alpha i+\beta j,
\gamma i+\delta j}} & \mbox{if} & d \ \mbox{is odd} \\ \\ 
(-1)^{(i+1)(j+1)}e^{2i\theta_{\alpha i+\beta j,
\gamma i+\delta j}} & \mbox{if} & d 
\ \mbox{is even.} \end{array} \right. \end{equation} 

\noindent {\sl where $\alpha, \beta, \gamma, \delta$ are integers modulo $d$ such that 
$\alpha \delta - \beta \gamma = \pm 1$.} 

\ 

\noindent The fiducial 14a (which is in the same field as 14b~\cite{ACFW}) does 
not appear in the tables because 
its higher dimensional cousin is not available at the moment.\footnote{Andrew Scott 
kindly produced the fiducials 120c and 195bcd when we asked for them.} With 
this exception the observations have been made 
starting from every SIC in dimension $4 \leq d \leq 15$. 

The integers occurring in the second observation can be collected into a 
matrix $M$, 
\begin{equation} M = \left( \begin{array}{cc} \alpha & \beta \\ \gamma & \delta 
\end{array} \right) \ , \hspace{8mm} \det{M} = \pm 1 \ {\rm mod} \ d \ . 
\label{matrisenM} \end{equation}

\noindent (The arithmetic is modulo $d$ also if $d$ is even.) In general this is 
an ESL matrix belonging to some coset of the symmetry group of the SIC. One 
can change the coset by 
choosing different SICs belonging to the same Clifford orbit.

The observations hold as stated 
only if the SIC fiducials are centred. If a displaced fiducial is used to calculate 
the overlaps then $(d-2)$th roots of unity appear in $e^{i\Theta_{di,dj}}$, and 
$d$th roots of unity in $e^{i\Theta_{(d-2)i,(d-2)j}}$. If the dimension $N$ is 
divisible by 3, as will always be the case from the third rung of the ladders and 
upwards, there are three SIC vectors in the same Zauner subspace. Unless 
one chooses the right one, roots of unity will again complicate the observations. 
It is natural to expect that the `right ones' can be taken to be strongly centred, 
but in those cases where an exact solution is missing we are unable to check 
this. Instead we refer to `suitably chosen' SIC vectors. 

With this understanding the observations hold for every adjacent pair of SICs in 
the columns of Tables \ref{tab:fiducials3} and \ref{tab:fiducials}. They motivate 
a formal definition:

\

\noindent {\bf Definition}. {\sl Pairs of SICs for which fiducial vectors can be 
chosen so that the two observations 
hold are aligned. The higher dimensional member of an aligned pair is called 
an aligned SIC.} 

\

\noindent There may well be logical dependencies among the two observations. Indeed, 
as we proceed we will find some evidence that this is so. Hence a more economical 
statement of the definition should be possible. 

Based on the fact that the two observations hold in every case we have looked at, 
we make the following conjecture. 

\

\noindent {\bf Conjecture}. Every $d$-dimensional Weyl--Heisenberg SIC has a 
corresponding aligned SIC in dimension $d(d-2)$.

\

\noindent It is worth noting that this conjecture is both stronger and weaker than 
the simple conjecture that SICs exist in every dimension. It posits significantly 
more structure on the problem, and is in that sense stronger. But it allows for the 
possibility that some dimensions might not contain SICs, or be otherwise sporadic, 
while still positing the existence of infinite families.  It also suggests a natural 
line of attack using inductive reasoning, though our own efforts in this direction 
have not yet been successful. But note also that the theorems in 
Sections~\ref{sec:entangle}--\ref{sec:sym} do 
not depend on the conjecture. They only depend on the (non-empty) definition. 

\vspace{1cm}

\section{Equiangular tight frames}\label{sec:equi}

\vspace{5mm}

\noindent The previous section clearly draws attention to two special subsets of 
vectors in an $N = d(d-2)$ dimensional SIC, namely 
\begin{equation}  \left\{ |\Psi_{(d-2)i,(d-2)j}\rangle \right\}_{i,j=0}^{d-1} 
\hspace{5mm} \mbox{and} \hspace{5mm} 
\{ |\Psi_{di,dj}\rangle \}_{i,j=0}^{d-3} \ . \label{etf} \end{equation}

\noindent The mutual overlaps within these subsets are very special numbers. What 
geometrical properties do these sets of vectors have?

A symmetric rank 1 POVM, also known as an equiangular tight frame (ETF), 
is a collection of $n$ unit vectors in ${\bf C}^m$ such that they resolve the 
identity, 
\begin{equation} \sum_{I=1}^n |\psi_I\rangle \langle \psi_I| = \frac{n}{m} 
{\mathbbm 1} \ , \label{povmcond} \end{equation}

\noindent and such that the absolute values $|\langle \psi_I|\psi_J \rangle |$ 
are equal whenever $I \neq J$. (We denote the dimension by $m$ since we cannot 
use $d$, for a reason that will soon be evident.) It is easy to show that $n$ 
cannot be smaller than $m$, and it cannot be larger than $m^2$~\cite{BF}. A minimal 
ETF is an orthonormal basis and a maximal ETF is a SIC, but there are many interesting 
intermediate cases~\cite{Mixon}. Because 
the overlaps $\langle \psi_I|\psi_J\rangle$ have constant absolute values it is easy 
to show---by squaring and taking the trace---that we must have 
\begin{equation} |\mbox{overlap}|^2 = \frac{n-m}{m(n-1)} \ . \end{equation}

\noindent Now let us fix an arbitrary integer $d > 3$, and ask for solutions of 
the Diophantine equation 
\begin{equation} \frac{n-m}{m(n-1)} = \frac{1}{d(d-2)+1} = 
\frac{1}{(d-1)^2} \ . \label{n} \end{equation}

\noindent There are typically many solutions. We are interested in four of them, namely 
\begin{equation} (m,n) = \left\{ \begin{array}{lll} \left( d(d-2),d^2(d-2)^2\right) 
& \ & {\rm SIC} \\ \\ 
\left( \frac{d(d-1)}{2}, d^2\right) & \ & {\rm ETF}_1 \\ \\  
\left( \frac{(d-1)(d-2)}{2}, (d-2)^2\right) & \ & {\rm ETF}_2  \\ \\ 
(d-1,d) & \ & {\rm ETF}_3 \ . \end{array} \right. 
\label{ETFs} \end{equation}

\noindent The first is that of a SIC in dimension $N = d(d-2)$. The fourth is a 
regular simplex in dimension $d$. The second and third solutions have just the right 
number of vectors to be identified with the equiangular subsets of the $N$-dimensional 
SIC that we identified above. 

The point here is that we have checked numerically, with a precision of 120 digits, that in 
each of the 19 aligned SICs listed in Section~\ref{sec:equi} the $d^2$ vectors in the first subset 
identified in (\ref{etf}) are linearly dependent and belong to a subspace of 
dimension $d(d-1)/2$. Similarly, the $(d-2)^2$ vectors in the second subset of 
(\ref{etf}) are linearly dependent and belong to a subspace of dimension $(d-1)(d-2)/2$. 
Hence they form smaller 
equiangular tight frames embedded in the aligned SIC. In the sequel, we will prove 
that this must happen in all aligned SICs (although the case of even $d$ is postponed 
to a later publication). We will also identify special aligned SICs which 
contain embedded $(d-1)$-dimensional simplices. 

\vspace{1cm}

\section{Entanglement properties of SIC vectors}\label{sec:entangle}

\vspace{5mm}

\noindent We now restrict the dimension of Hilbert space to be odd, for the pragmatic 
reason that then the Weyl--Heisenberg group defines a preferred tensor product 
decomposition ${\mathbb C}^{d(d-2)} = {\mathbb C}^d\otimes {\mathbb C}^{d-2}$. As a result
every vector in ${\mathbb C}^{d(d-2)}$ can be described in the language of entanglement 
theory. In particular we will find the Schmidt decomposition very useful. Although this 
language is familiar to every quantum information scientist, we recall the basic facts 
that we need. Better explanations can be found elsewhere~\cite{Ekert}.

Suppose that ${\mathbb C}^N = 
{\mathbb C}^{n_1}\otimes {\mathbb C}^{n_2}$ where $n_1 \geq n_2$. There will be local 
operators affecting only one of the factors of the Hilbert space. Given a 
pure state vector $|\Psi \rangle$ in the large Hilbert space, we define a reduced 
state $\rho_1$, which is a density matrix acting on ${\mathbb C}^{n_1}$, by the 
requirement that for all operators of the form $A_1\otimes {\mathbbm 1}$ there 
holds  
\begin{equation} 
\mbox{Tr} |\Psi \rangle \langle \Psi |(A_1\otimes {\mathbbm 1}) 
= \mbox{Tr}_1 \rho_1 A_1 \ , \end{equation}

\noindent where $\mbox{Tr}_1$ denotes the trace over matrices acting on ${\mathbb C}^{n_1}$. 
This is enough to define $\rho_1$. One can explicitly write 
\begin{equation} \rho_1 = \mbox{Tr}_2|\Psi \rangle \langle \Psi | \ , \end{equation}

\noindent where $\mbox{Tr}_2$ denotes the partial trace over the second factor. 
The reduced state $\rho_2$ is defined similarly, using a partial trace over the first 
factor. Although the state we start out from is pure (defines a 
one-dimensional projector), the reduced state $\rho_1$ is typically a convex mixture 
of more than one pure state acting on ${\mathbb C}^{n_1}$. 
Generically it will have $n_2$ non-vanishing eigenvalues. A comfortable theorem 
says that the spectra of $\rho_1$ and $\rho_2$ are identical, except for additional 
zero eigenvalues in the larger dimension. The eigenvalues $\lambda_k$ of the reduced 
density matrices are called Schmidt coefficients, and they completely determine the 
entanglement properties of a pure state $|\Psi\rangle $ in dimension $N = n_1n_2$. 
Indeed, given any such pure state $|\Psi\rangle$ one can always adapt the orthonormal 
bases $\{ |e_k\rangle \}_{k=0}^{n_1-1}$ and $\{ |f_k\rangle \}_{k=0}^{n_2-1}$ in the 
factors, such that $|\Psi \rangle$ is given by the single sum
\begin{equation} |\Psi \rangle = \sum_{k=0}^{n_2-1}\sqrt{\lambda_k}|e_k\rangle |f_k\rangle \ . 
\end{equation}

\noindent This is called the Schmidt decomposition of the state, and the coefficients 
in this expansion are the positive square roots of the Schmidt coefficients. Practical 
computation of the Schmidt decomposition follows by noting that the singular value 
decomposition of the $n_1\times n_2$ matrix whose entries are the components of 
$|\Psi \rangle$ gives the same information.     

We can now ask: what are the entanglement properties of a SIC vector in dimension 
$N = d(d-2)$? For generic pure states one expects $d-2$ different, and non-vanishing, 
Schmidt coefficients, but we will prove that the vectors in an aligned SIC are 
highly non-generic in this regard.

At the outset we consider dimension $N = n_1n_2$, where $n_1$ and $n_2$ are 
relatively prime and odd. We use the fact that the Weyl--Heisenberg group is a unitary 
operator basis, and then the group isomorphism provided by the Chinese remainder 
theorem, to conclude for any vector $|\Psi\rangle \in {\mathbb C}^N$ that 
\begin{eqnarray} |\Psi \rangle \langle \Psi | &=& \frac{1}{N}\sum_{i,j=0}^{N-1} 
D_{-i,-j}^{(N)}\langle \Psi |D^{(N)}_{i,j}|\Psi \rangle = \nonumber \\  \\ 
&=& \frac{1}{n_1n_2}\sum_{i_1,j_1=0}^{n_1}\sum_{i_2,j_2=0}^{n_2}
D^{(n_1)}_{-i_1,-n_2^{-1}j_1}\otimes D^{(n_2)}_{-i_2,-n_1^{-1}j_2} 
\langle \Psi | D_{i,j}^{(N)} | \Psi \rangle \ , \nonumber \end{eqnarray}

\noindent where applying the Chinese remainder theorem (see Appendix~\ref{sec:CRT}) allows us to express 
\begin{equation} \langle \Psi | D^{(N)}_{i,j} | \Psi \rangle = 
\langle \Psi | D^{(N)}_{i_1n_2n_2^{-1} + i_2n_1n_1^{-1},\ j_1n_2n_2^{-1}+j_2n_1n_1^{-1}} | 
\Psi \rangle \ .  \end{equation}

\noindent If we now take the partial trace over, say, the first factor only the 
terms with $i_1 = j_1 = 0$ contribute. In this way we obtain the reduced density 
matrix  
\begin{equation} \rho^{(n_2)} = 
\mbox{Tr}_{n_1}|\Psi \rangle \langle \Psi | = \frac{1}{n_2}
\sum_{i_2,j_2=0}^{n_2-1}D_{-i_2,-j_2}^{(n_2)} 
\langle \Psi |D^{(N)}_{i_2n_1n_1^{-1},\ j_2n_1}|\Psi \rangle \ . \label{red} \end{equation}

\noindent One summation index was shifted, which is allowed. 

Now we specialize to the case of interest, namely 
\begin{equation} n_1 = d \ , \hspace{8mm} n_2 = d-2 \ , 
\hspace{8mm} n_1^{-1} = n_2^{-1} = \frac{d-1}{2} \equiv \kappa \ , \end{equation}

\noindent and to the case that $|\Psi\rangle$ is a vector in an aligned SIC. 
We drop the subscripts on the indices---which are no longer needed since they are 
summation indices only---and conclude from the above that 
\begin{equation} \rho^{(d-2)} = \frac{1}{d-2}\sum_{i,j=0}^{d-3}D^{(d-2)}_{-i,-j}
\langle \Psi |D^{(N)}_{id\kappa , jd}|\Psi \rangle \ . \label{prelspar} \end{equation}

\noindent We are now ready to prove our first theorem. The parity operator that 
occurs in its statement is defined in Appendix~\ref{sec:parity}. 

\

\begin{theorem}\label{thm:1}
{\sl If $d$ is odd and if $|\Psi_0\rangle$ is a 
suitably chosen SIC vector in an aligned SIC in dimension $d(d-2)$, the density 
matrix reduced to dimension $d-2$ is}
\begin{equation} \rho_{\bf 0}^{(d-2)} \equiv \mbox{Tr}_{d}|\Psi_{\bf 0} \rangle \langle 
\Psi_{\bf 0} | = \frac{1}{d-1}( {\mathbbm 1}_{d-2} + P^{(d-2)} ) \ , \end{equation}

\noindent {\sl where $P^{(d-2)}$ is the parity operator in dimension $d-2$. 
Hence $\rho^{(d-2)}_{\bf 0}$ is proportional to a projector from ${\mathbb C}^{d-2}$ 
onto a subspace of dimension $(d-1)/2$.}
\end{theorem}
\

\noindent {\it Proof}: Recalling that we defined $e^{i\Theta_{0,0}} = 1$ we 
rewrite Eq. (\ref{prelspar}) as   
\begin{equation} \mbox{Tr}_{d}|\Psi_{\bf 0} \rangle \langle \Psi_{\bf 0} | = \frac{1}{d-2}
\left( \left(1-\frac{1}{d-1}\right) {\mathbbm 1} + \frac{1}{d-1}
\sum_{i,j=0}^{d-3}D_{-i,-j}^{(d-2)} e^{i\Theta_{d\kappa i,dj}}\right)\ . 
\label{proof1} \end{equation}

\noindent The definition of an aligned SIC implies that we can choose the fiducial 
so that 
\begin{equation} e^{i\Theta_{d\kappa i,dj}} = 1 \ . \end{equation}

\noindent Equation (\ref{proof1}) then becomes
\begin{equation} \rho_{\bf 0}^{(d-2)} = \frac{1}{d-1} \left( {\mathbbm 1} + \frac{1}{d-2}
\sum_{i,j=0}^{d-3}D_{-i,-j}^{(d-2)}\right) = \frac{1}{d-1}( {\mathbbm 1}_{d-2} + P^{(d-2)} ) \ , 
\end{equation} 

\noindent where Eq. (\ref{parity}) for the parity operator was used in the last step. 
In dimension $d-2$ the operator $({\mathbbm 1} + P)/2$ is a projection operator of rank 
$(d-1)/2$, which gives the final part of the statement. $\Box$ 

\

\noindent Thus we find only $(d-1)/2$ non-vanishing Schmidt coefficients, and they are 
all equal. 
Indeed 
the entanglement properties of a vector belonging to an aligned SIC are very special. 

The theorem applies only to aligned SICs, such as 15d and 195abcd. A calculation shows 
that the non-aligned fiducials 15abc have non-degenerate Schmidt coefficients, as 
expected for generic vectors. (Compare Table \ref{tab:fiducials3}.) On the other hand the 
restriction to special choices of SIC vectors can be removed, except that one then 
encounters displaced parity operators on the right hand side. The proof 
simplifies considerably if we choose the fiducials suitably. 

The next task is to find the state reduced to dimension $d$. From entanglement theory 
we know that the spectra of Tr$_d|\Psi_0\rangle \langle \Psi_0|$ and 
Tr$_{d-2}|\Psi_0\rangle \langle \Psi_0|$ coincide. However, the precise 
mechanism that allows this to happen is worth studying because it depends on 
the details of our definition of aligned SICs. This will show that the two observations 
we made are in fact related. 

The preliminary steps are the same as before. In Eq. (\ref{red}), set 
$(n_1,n_2) = (d-2,d)$, and rewrite 
\begin{eqnarray} \mbox{Tr}_{d-2}|\Psi_{\bf 0} \rangle \langle \Psi_{\bf 0} | &=& \frac{1}{d}
\left( {\mathbbm 1} + \frac{1}{d-1}
\sum_{i,j\neq (0,0)}^{d-1}D_{-i,-j}^{(d)} e^{i\Theta_{(d-2)\kappa i,(d-2)j}}\right) 
= \nonumber \\ \label{proof2} \\ 
&=& \frac{1}{d}
\left( {\mathbbm 1} + \frac{1}{d-1}
\sum_{i,j\neq (0,0)}^{d-1}D_{2i,-j}^{(d)} e^{i\Theta_{(d-2) i,(d-2)j}} \right) 
\nonumber \end{eqnarray}

\noindent We are now ready to bring in the squared overlap phases in dimension $d$ 
by applying the full definition of an aligned SIC. 

\

\begin{theorem}\label{thm:2}
{\sl If $d$ is odd and if $|\Psi_0\rangle$ is a 
suitable SIC vector in an aligned SIC in dimension $d(d-2)/2$, the density 
matrix reduced to dimension $d$ is}
\begin{equation} \rho_{\bf 0}^{(d)} \equiv \mbox{Tr}_{d-2}|\Psi_{\bf 0} \rangle \langle 
\Psi_{\bf 0} | = 
\frac{1}{d-1}( {\mathbbm 1}_d - P^{(d)}_\theta ) \ , \end{equation}

\noindent {\sl where $P^{(d)}_\theta$ is a generalized parity operator in 
dimension $d$. Hence $\rho^{(d)}_0$ is proportional to a projector from 
${\mathbb C}^{d}$ onto a subspace of dimension $(d-1)/2$.}
\end{theorem}
\

\noindent {\it Proof}: Applying the definition of an aligned SIC to Eq. 
(\ref{proof2}) we obtain 
\begin{eqnarray}  \rho_{\bf 0}^{(d)} &=& \frac{1}{d}
\left( \left(1+\frac{1}{d-1}\right) {\mathbbm 1} - \frac{1}{d-1}
\sum_{i,j=0}^{d-1}D_{2i,-j}^{(d)} e^{2i\theta_{\alpha i + \beta j, \gamma i + 
\delta j}}\right) = 
\nonumber \\ \\ 
&=& \frac{1}{d-1}
\left( {\mathbbm 1} - \frac{1}{d}
\sum_{i,j=0}^{d-1}D^{(d)}_{-i,-j} e^{2i\theta_{-2^{-1}\alpha i + \beta j,-2^{-1}\gamma i + 
\delta j}} \right) \ . 
\nonumber \end{eqnarray}

\noindent We relabeled the summation index and introduced the multiplicative 
inverse of $2$ modulo $d$. Making use of Eq. (\ref{matrisenM}) 
\begin{equation}  \rho_{\bf 0}^{(d)} = \frac{1}{d-1}\left( {\mathbbm 1} - \frac{1}{d}
\sum_{\bf p}D^{(d)}_{-{\bf p}} e^{2i\theta_{M'{\bf p}}} \right) \ , \end{equation}

\noindent where the $GL(2,{\mathbb Z}_d)$ matrix $M'$ obeys 
$\det{M'^{-1}} = \pm 2$. We now appeal to a result from ref.~\cite{Dardo}, which says 
that, under the conditions stated, the generalized parity operator 
\begin{equation} P_\theta = \frac{1}{d}
\sum_{\bf p}D_{-{\bf p}} e^{2i\theta_{M'{\bf p}}} \label{genparity} \end{equation} 

\noindent obeys $P_\theta^2 = {\mathbbm 1}$ and has $(d+1)/2$ eigenvalues equal to $+1$ 
and $(d-1)/2$ eigenvalues equal to $-1$. $\Box$

\

\noindent Concerning the result from ref.~\cite{Dardo} we observe that it is 
a consequence of a key property of SICs, that they form projective 2-designs. 
This goes some way towards explaining why squared overlap phases play a role. 
See ref.~\cite{Belovs} for a review of projective $t$-designs.  

Again the restriction to special choices of fiducials can be dropped at 
the expense of complicating the statement of the theorem a little, and significantly 
complicating the direct proof. In Section~\ref{sec:embed} we will formulate a geometrical theorem 
where this restriction is dropped. 

\vspace{1cm}

\section{Mutually unbiased bases}\label{sec:MUB}

\vspace{5mm}

\noindent The appearance of the parity operator $P$ in the preceding section allows us 
to give a resolution of the long-standing question of how to relate SICs to mutually 
unbiased bases (MUB) in prime dimensions. By definition a complete set of MUB in 
dimension $p$ is a collection of $p+1$ orthonormal bases such that every overlap between 
vectors in different bases has absolute value squared equal to $1/p$~\cite{Ivanovic}. 
This definition, 
like the definition of a SIC, has its origin in quantum state tomography, and MUB 
have found a number of interesting applications over the years. Complete sets of MUB 
do exist in all dimensions equal to a power of a prime number~\cite{Wootters2}, and 
if the dimension $p$ is a prime number they arise as eigenbases of the $p+1$ cyclic 
subgroups of the Weyl--Heisenberg group. (If the dimension is equal to a higher power 
of a prime number a multipartite Heisenberg group appears. In non-prime power dimensions 
complete sets of MUB may well not exist, and if they do they are unrelated to the 
Heisenberg groups~\cite{Zauner, ACW}.) Given this group theoretical connection one 
expects to find a tight geometrical connection between MUB and SICs in prime dimensional 
Hilbert spaces. This is indeed so in the very special case of $d = 3$, which was cleared 
up in 1844~\cite{Hesse}. When $d > 3$ it has to be kept in mind that MUB are based on 
cyclotomic fields, while SICs are two steps 
beyond that since ray class fields over real quadratic fields come in. 
Although a loose connection between SICs and MUB in prime dimensions 
exists~\cite{ADF}, the details have remained elusive. 

We can now offer an answer to this question, because our Theorem \ref{thm:1} provides us 
with the means to construct a complete set of MUB in dimension $p = d-2$ (assumed to 
be a prime number) from an aligned SIC in dimension $N = d(d-2)$~\cite{Ivanovic}. In 
fact, given Wootters' elegant construction of complete sets of MUB in prime dimensions~\cite{Wootters}, this result follows trivially from the above, but the details 
are worth spelling out. The starting point is the observation that in prime dimension 
the vectors labeling the displacement operators form a true vector space. This is so 
because the set of integers modulo a prime number 
form a finite field. This vector space can be regarded as a finite 
affine plane consisting of $p^2$ points and $p(p+1)$ lines containing $p$ points 
each. The lines are given by the equation 
\begin{equation} j = zi + a \ , \label{za} \end{equation}

\noindent where $i,j,a$ are integers modulo $p$ while $z$ can also take the formal 
value $\infty$, corresponding to a set of `vertical' lines~\cite{ADF}. Thus a line 
is given by fixing the pair $(z,a)$. Next, consider the $p^2$ displaced 
parity operators
\begin{equation} P_{i,j} = D_{i,j}PD_{-i,-j} \ . \label{partitydis} \end{equation}

\noindent They are renamed as phase point operators, and associated with the $p^2$ 
points of the affine plane. We also need operators associated with the $p(p+1)$ lines 
of the affine plane. A key fact proved by Wootters is that the operators
\begin{equation} W^{(z,a)} = \frac{1}{p}\sum_{\rm line} P_{i,j} \label{gamleW} 
\end{equation}

\noindent are one-dimensional projectors projecting to the vectors in a complete 
set of MUB. The sum goes over all $i,j$ consistent with Eq. (\ref{za}) for some 
given $z,a$. The construction needs the combinatorics of the affine plane to work, 
which is certainly available when $p$ is prime.  

We now have:

\

\begin{theorem}\label{thm:3}
{\sl If $p=d-2$ is an odd prime then a complete set 
of MUB in dimension $p$ can be obtained by taking affine combinations of projectors 
to the vectors in an aligned SIC in dimension $d(d-2)$, and then performing a partial 
trace.}
\end{theorem}
\

\noindent {\it Proof}: By Theorem \ref{thm:1} and the properties of the partial trace  
\begin{equation} \mbox{Tr}_d \left( {\mathbbm 1}_d\otimes D^{(d-2)}_{i,j}\right)
|\Psi_{\bf 0}\rangle \langle \Psi_{\bf 0}|\left( {\mathbbm 1}_d\otimes D^{(d-2)}_{-i,-j}
\right) = \frac{1}{d-1}\left( {\mathbbm 1}_{d-2} + 
P_{i,j}\right) \ , \end{equation}

\noindent where we used definition (\ref{partitydis}) for the displaced parity operators 
in dimension $d-2$. The construction uses the $p^2 = (d-2)^2$ SIC vectors 
\begin{equation} |\Psi_{di,dj}\rangle = {\mathbbm 1}_d\otimes D_{di,j}^{(d-2)}|\Psi_{\bf 0}\rangle 
\ . \end{equation} 

\noindent Using Wootters' formula (\ref{gamleW}), and the linearity of the 
trace, we immediately obtain
\begin{equation} W^{(z,a)} = \mbox{Tr}_d\left[ \frac{d-1}{d-2}\sum_{\rm line} 
|\Psi_{di,dj}\rangle \langle \Psi_{di,dj}| - \frac{1}{d}{\mathbbm 1}_N\right] 
\ . \end{equation}

\

\noindent By construction the $(p+1)p$ operators $W^{(z,a)}$ project to the vectors 
in a complete set of MUB. $\Box$

\

\noindent Hence we have a firm relation between MUB in dimension $p$ and SICs in 
dimension $(p+2)p$. Unfortunately we do not have a way to go from SICs in dimension 
$d$ to SICs in dimension $d(d-2)$, nor are we close to having this, but if we had 
we would have a firm relation between MUB in dimension $p$ and SICs in 
dimension $p+2$. 

\vspace{1cm}

\section{The embedding of the equiangular tight frames}\label{sec:embed}

\vspace{5mm}

\noindent We are now ready to prove (for odd $d$) that the equiangular tight 
frames observed in Section~\ref{sec:equi} have to appear in every aligned SIC. 
Because the Weyl-Heisenberg group is an operator basis Schur's lemma 
implies, for any operator $A$, that 
\begin{equation} \frac{1}{N}\sum_{\bf p}D_{\bf p}AD^{\dagger}_{\bf p} = {\mathbbm 1}_N 
\mbox {Tr}A \ . \end{equation}

\noindent Now suppose the dimension is composite, $N = n_1n_2$, and assume that the 
factors are relatively prime and odd. Then Chinese remaindering can be applied, and 
one can show that   
\begin{equation} \frac{1}{n_1}\sum_{{\bf p}_1}(D^{(n_1)}_{{\bf p}_1} 
\otimes {\mathbbm 1}_{n_2}) A (D^{(n_1)}_{-{\bf p}_1}\otimes {\mathbbm 1}_{n_2}) = 
{\mathbbm 1}_{n_1}\otimes \mbox{Tr}_{n_1}A \ . \label{parttrace0} \end{equation}

\noindent We have `isotropized' one factor of the tensor product, and a partial 
trace appears on the other. A similar equation, with the role of the factors 
interchanged, will also be used below. 

We now specialize to the case $n_1 = d$, $n_2 = d-2$, and $A = |\Psi_{\bf 0}\rangle 
\langle \Psi_{\bf 0}|$, where $|\Psi_{\bf 0}\rangle$ is a suitably chosen 
SIC vector aligned 
with a SIC vector in dimension $d$. Then Theorems \ref{thm:1} and \ref{thm:2} give us information 
about the partial trace that appears on the right hand side. On the other hand, 
the left hand side has an interesting interpretation. Indeed, we can consider 
the two operators 
\begin{eqnarray} \Pi_1 &\equiv & \frac{d-1}{2d}\sum_{i,j=0}^{d-1}|\Psi_{(d-2)i,(d-2)j}\rangle 
\langle \Psi_{(d-2)i,(d-2)j}| =  \nonumber  \\ \label{def1} \\ 
&=& \frac{d-1}{2} \frac{1}{d}\sum_{{\bf p}_1}
(D^{(d)}_{{\bf p}_1} \otimes {\mathbbm 1}_{d-2}) |\Psi_{\bf 0}\rangle \langle 
\Psi_{\bf 0}| (D^{(d)}_{-{\bf p}_1}\otimes {\mathbbm 1}_{d-2}) \nonumber \end{eqnarray}
\begin{eqnarray} \Pi_2 &\equiv & \frac{d-1}{2(d-2)}\sum_{i,j=0}^{d-3}|\Psi_{di,dj}\rangle 
\langle \Psi_{di,dj}| =  \nonumber \\ \label{def2} \\ 
&=& \frac{d-1}{2} \frac{1}{d-2}\sum_{{\bf p}_2}
({\mathbbm 1}_{d}\otimes D^{(d-2)}_{{\bf p}_2} ) |\Psi_{\bf 0}\rangle \langle 
\Psi_{\bf 0}| ({\mathbbm 1}_{d}\otimes D^{(d-2)}_{-{\bf p}_2}) \ . \nonumber \end{eqnarray} 

\noindent The idea behind the next theorem is that these operators are projectors, 
and can be substituted for the unit operator in the POVM condition (\ref{povmcond}) 
provided we restrict ourselves to the subspaces of ${\mathbb C}^N$ to which these 
operators project.  

\

\begin{theorem}\label{thm:4}
{\sl If $d$ is odd, then every aligned SIC in 
dimension $d(d-2)$ contains two multiplets of smaller equiangular tight frames 
embedded in it. Each individual SIC vector in an aligned SIC belongs to an 
equiangular tight frame of $d^2$ vectors spanning a subspace of dimension 
$d(d-1)/2$, and another consisting of $(d-2)^2$ vectors spanning a subspace of 
dimension $(d-1)(d-2)/2$.}
\end{theorem}

\

\noindent {\it Proof}: Combining the definitions (\ref{def1}) and (\ref{def2}), 
Eq. (\ref{parttrace0}), and Theorems \ref{thm:1} and \ref{thm:2}, gives immediately that 
\begin{equation} \Pi_1 = {\mathbbm 1}_d \otimes \frac{1}{2}({\mathbbm 1}_{d-2} + P^{(d-2)}) 
\ \ \end{equation}
\begin{equation} \Pi_2 = \frac{1}{2}({\mathbbm 1}_{d} - P^{(d)}_{\theta}) \otimes 
{\mathbbm 1}_{d-2} \ . \end{equation}

\noindent It follows that $\Pi_1$ and $\Pi_2$ are projectors, to subspaces of 
dimension $d(d-1)/2$ and $(d-1)(d-2)/2$, respectively. To see that the support 
of $\Pi_1$ contains $d^2$ equiangular SIC vectors one performs the calculation 
\begin{eqnarray} \langle \Psi_{(d-2)i,(d-2)j}|\Pi_1|\Psi_{(d-2)i,(d-2)j}\rangle = 
\hspace{20mm} \nonumber \\ \\ 
\hspace{20mm} = \mbox{Tr} \Pi_1|\Psi_{(d-2)i,(d-2)j}\rangle \langle 
\Psi_{(d-2)i,(d-2)j}| = 1 \ , \nonumber \end{eqnarray}

\noindent and similarly for $\Pi_2$. The fiducial $|\Psi_{\bf 0}\rangle$ belongs 
to both subspaces. Conjugating with the Weyl--Heisenberg group one finds that the 
subspace defined by the projector $\Pi_1$ belongs to an orbit of $(d-2)^2$ subspaces 
each containing an ETF of type $\left( d(d-1)/2, d^2\right)$, and similarly 
for $\Pi_2$. $\Box$

\

The projectors $\Pi_1$ and $\Pi_2$, and the Gram matrices of the 
resulting ETFs, are constructed entirely out of numbers present in the 
$d$-dimensional SIC and of suitable roots of unity. Waldron~\cite{Waldron} 
and Goyeneche have already noted that given a SIC in dimension $d$ 
one can always construct the Gram matrices corresponding to equiangular tight frames 
of the types we have here found to be embedded in the aligned $d(d-2)$-dimensional 
SICs. This result is valid regardless of whether $d$ is odd and even.
A version of 
Theorem \ref{thm:4} that holds for arbitrary $d$ is in fact known, but we postpone its 
presentation to a companion paper. 

In Eq. (\ref{ETFs}) we also raised the possibility that a regular $(d-1)$-dimensional 
simplex can be embedded in a $d(d-2)$-dimensional SIC. This happens in three of 
our examples, namely 8b, 35j, and 120c, and is connected (via our definition of 
aligned SICs) to the fact that $d-1$ real overlap phases $e^{i\theta_{i,j}}$ 
occur in the relevant $d$-dimensional SICs 4a, 7b, and 12b, all of which have an 
extra anti-unitary symmetry beyond the 
Zauner symmetry. This is not a property that is inherited on higher rungs of the 
ladder though; 8b has only 3 real phases, and 35j only 30 real phases. 

The embedding of lower dimensional ETFs in the SIC means that non-trivial linear 
dependencies are present among the vectors of the latter. The general question 
under what conditions sets of vectors in Weyl--Heisenberg orbits can be linearly 
dependent has been studied~\cite{Hoan, Diogenes}, and it is known that linear 
dependencies do occur, in such orbits, whenever the order of their symmetry group 
fails to be coprime with the dimension. Some of the linear dependencies that we 
report here are not covered by these results. 

\vspace{1cm}

\section{Symmetries}\label{sec:sym}

\vspace{5mm}

\noindent A striking feature of Tables \ref{tab:fiducials3} and \ref{tab:fiducials} is 
that the order of the intrinsic symmetry group of the SICs increases with 
a factor of two for each rung of the ladder. In fact several of the numerical fiducials 
in these high dimensions were found because Scott and Grassl~\cite{Scott, Andrew} 
conjectured the presence of an extra symmetry of order 2 (beyond the order 3 Zauner 
symmetry), given by the symplectic matrix 
\begin{equation} F_b = \left( \begin{array}{cc} 1-d & 0 \\ 0 & 1-d \end{array} \right) 
\in SL(2,{\mathbb Z}_N) \ . \end{equation}

\noindent In the standard representation that we use~\cite{Marcus} an easy calculation 
gives, after Chinese remaindering according to Eq. (\ref{Cliffordsplit}), that the 
corresponding unitary operator is 
\begin{equation} U_b = {\mathbbm 1}\otimes P \ , \end{equation}

\noindent where $P$ is the parity operator in dimension $d-2$. It is easy to prove that 
this symmetry has to be there. 

\

\begin{theorem}\label{thm:5}
{\sl An aligned SIC in an odd dimension 
is invariant under $U_b$.}  
\end{theorem}
\

\noindent {\it Proof}: Let $|\Psi_0\rangle$ be a strongly centred   
SIC fiducial. Then Theorem \ref{thm:1} states that the reduced density matrix is 
\begin{equation} \rho_2 = \mbox{Tr}_1|\Psi_{\bf 0}\rangle \langle \Psi_{\bf 0}| = 
\frac{1}{d-1}({\mathbbm 1} + P) \ . \end{equation}

\noindent The Schmidt decomposition~\cite{Ekert} of such a state is   
\begin{equation} |\Psi_{\bf 0}\rangle = \sqrt{\frac{2}{d-1}}\sum_{k=1}^{\frac{d-1}{2}}|e_k\rangle 
|f_k\rangle  \ . \end{equation}

\noindent Moreover $\rho_2$ and $U_b = {\mathbbm 1}\otimes P$ are diagonal in the Schmidt basis, 
and $U_b$ manifestly leaves $|\Psi_0\rangle$ invariant. Being a member of the Clifford 
group it will permute the remaining SIC vectors among themselves. $\Box$

\

\noindent A similar argument fails on the left hand factor. The generalized parity 
operator can be used to construct an operator that leaves $|\Psi_{\bf 0}\rangle$ 
invariant, but since it is not a member of the Clifford group the last line in the 
proof fails. This is also the reason why, in Section~\ref{sec:MUB}, we were able to connect 
aligned SICs to mutually unbiased bases in dimension $d-2$, but not to MUB in 
dimension $d$. 

There is more to say about symmetries and dimension towers, and we hope to come 
back to these issues in a later publication. 

\vspace{1cm}

\section{Conclusions}\label{sec:conc}

\vspace{5mm}

\noindent The number theoretical connections between SICs in dimension $d$ 
and dimension $d(d-2)$ manifest themselves very explicitly in the case of 
aligned SICs. The number field needed to construct the former is a subfield 
of that needed to construct the latter~\cite{AFMY, AFMY2}. Gary McConnell has noted 
that it can happen that some of the overlap phases in dimension $d(d-2)$ actually 
belong to the subfield. We have explored a part of this pattern, and it 
enables us to make significant statements about the Hilbert space geometry 
of the relevant $d(d-2)$ dimensional SICs. Moreover we have collected evidence, 
in the form of 19 mostly numerical examples, suggesting that every SIC in 
dimension $d$ gives rise to a SIC in dimension $d(d-2)$ where this pattern 
occurs. The higher dimensional member of such a pair is said to be an 
aligned SIC, and we offered a precise definition of aligned SICs.

In this paper we concentrated on the case of odd dimensions, in which case there 
is a canonical tensor product structure. Then the alignment manifests itself as very 
special entanglement properties (Theorems \ref{thm:1} and \ref{thm:2}). If $d-2 = p$ is an odd prime 
number a complete set of mutually unbiased bases in dimension $p$ can be derived 
from a higher dimensional SIC (Theorem \ref{thm:3}). We also proved that there are non-trivial 
equiangular tight frames embedded in the $d(d-2)$ dimensional aligned SICs 
(Theorem \ref{thm:4}). This property generalizes to even dimensions, as 
we will prove in a companion paper. Finally we proved that a conjectured extra 
symmetry is indeed always present in the aligned SICs (Theorem \ref{thm:5}). 

We stress that we have only scratched the surface of an intricate pattern. There is 
more to the story than just squared phases. Then, as we 
discussed in the introduction, there are other dimension towers to consider. 
The field inclusions organize the dimension towers into partially ordered sets 
with a very intricate structure. Moreover, very recently Grassl and Scott~\cite{GS} 
published the results of an investigation of the full sequence (\ref{seqD5}), 
corresponding to $D = 5$. They conjecture that the ray class SICs in these dimensions 
have a special symmetry that grows with $d$, and verify this 
conjecture by calculating an exact solution for $d = 124$ (!) as well as numerical 
solutions in dimensions $323$ and $844$ (!). Their approach is in a way complementary 
to ours, since we have not focussed on the ray class SICs exclusively. In fact, as 
our Figure \ref{fig:inclusions} may make clear, the full picture is likely to be even 
richer than what Figure \ref{fig:ladders} begins to suggest.

There is a hope that one can find a way to construct higher dimensional 
SICs starting from lower dimensional ones, and this hope has served as one 
of our motivations. There is also an over-riding question: What is the `mechanism' 
forcing certain algebraic number fields of great independent interest 
to manifest themselves in Hilbert space in the precise way they do? 
We are far from an answer, but we hope our results represent a small step 
forwards. 

\vspace{2cm}

\noindent \underline{Acknowledgements}: We thank Gary McConnell for allowing 
us to use some observations of his as our starting point. We would not have 
been able to do the work without his support, or without the continuous 
assistance of Andrew Scott (who deserves special thanks for finding the 
fiducials 195bcd and 120c for us). We also acknowledge Emily King for 
motivating discussions at a workshop arranged by the Hausdorff Institute, 
and Markus Grassl for spotting several mistakes in an almost final draft. 
This research was supported in part by the Australian Research Council via 
EQuS project number CE11001013. SF acknowledges support from an Australian 
Research Council Future Fellowship FT130101744. 
IB was not supported by the totally corrupt Swedish Research Council (VR). 

\newpage

\appendix
\vspace{1cm}

\section{Roots of unity}\label{sec:roots}

\vspace{5mm}

\noindent When it was first calculated the SIC in dimension 6 seemed to cement the 
idea that SICs are significantly more complex than mutually unbiased bases~\cite{Markus1}. 
However, on further 
reflection it will be seen that we were not really comparing apples to apples. The exact 
solutions for the known SICs are written in radicals.  If the number $e^{\frac{2\pi i}{n}}$ 
is written out in radicals the expression which results is also very complicated (except in 
special cases). Thus, using the techniques developed by Lagrange, Vandermonde, and 
Gauss~\cite{Tignol}, one finds that the primitive eleventh root of unity is 
\begin{eqnarray}
& \omega_{11} =-\frac{1}{10}+\left(\frac{1}{40} (-1+b_1)+\frac{1}{20} 
(1+b_1) b_2\right) b_3\nonumber \\
& \hspace{1 cm} +\left(\frac{1}{440} (-1+5 b_1)+\frac{1}{220}
(-5-b_1) b_2\right) b_3^2
\nonumber \\
&\hspace{ 1.1 cm} +\left(\frac{-1+4 b_1}{1210}+\frac{1}{605} (-2-2 b_1) b_2\right) b_3^3
\nonumber \\
&\hspace{ 1.2 cm} +\left(\frac{9+5
b_1}{13310}+\frac{(-45-3 b_1) b_2}{13310}\right) b_3^4
+\left(\frac{109-25 b_1}{585640}+\frac{(17+29 b_1) b_2}{58564}\right)
b_3^5
\\
&
+\left(\frac{29+505 b_1}{6442040}+\frac{(390+37 b_1) b_2}{1610510}\right) b_3^6
+\left(\frac{-1159-1519 b_1}{70862440}+\frac{(49-546
b_1) b_2}{17715610}\right) b_3^7
\nonumber \\
&
+\left(\frac{-619+7295 b_1}{779486840}+\frac{(2125+2129 b_1) b_2}{389743420}\right)
b_3^8
+\left(\frac{26459-14299 b_1}{8574355240}+\frac{(25829+10629 b_1) b_2}{4287177620}\right) b_3^9
\nonumber \end{eqnarray}
\noindent where
\begin{eqnarray}
b_1 &=& \sqrt{5} \ , \hspace{8mm} 
b_2 = \frac{i}{4}\sqrt{10-2b_1} \ , \\ 
b_3 &=& ( \tfrac{1}{4} (561671+29975 b_1)+(-24365+37620b_1)b_2)^{\frac{1}{10}} \ . 
\nonumber \end{eqnarray}

\noindent If this formula was used to calculate MUB in dimension 11 the complexity of 
the resulting expressions would be similar to the complexity of the expressions for the $d=11$ SICs given by 
Scott and Grassl~\cite{Scott}. On the other hand, using the transcendental function 
$f(z) = e^{2\pi i z}$, we find 
\begin{equation} \omega_{11} = f\left( \frac{1}{11}\right) \ . \end{equation}

\noindent Hilbert's 12th problem asks for a representation of the numbers needed to construct 
SICs analogous to the second description of the 11th root of unity. The suggestion is 
that SICs, if they could be seen through the 
right number theoretical glasses, are as simple as MUB in prime dimensions are. 

\vspace{1cm}

\section{The Weyl--Heisenberg and Clifford groups}\label{sec:WH}

\vspace{5mm}

\noindent We define the Weyl--Heisenberg group $H(d)$ in dimension $d$ to contain 
central elements represented by the phase factors~\cite{Marcus} 
\begin{equation} \tau = - e^{\frac{i\pi}{d}} \ , \hspace{10mm} \omega = \tau^2 
= e^{\frac{2\pi i}{d}} \ . \end{equation}

\noindent (Multiplication with the unit matrix is left understood whenever this cannot 
cause confusion.) If the dimension $d$ is odd, as we assume, then $(d+1)/2$ is an 
integer and there holds 
\begin{equation} \omega^{\frac{d+1}{2}} = \left( e^{\frac{\pi i}{d}}\right)^{d+1} 
= \tau \ . \end{equation}

\noindent Both $\tau$ and $\omega$ are $d$th roots of unity in this case. If $d$ is 
even some complications arise, and we postpone this case to a separate paper. Here 
we only wish to note the fact, evident from the introduction, that odd and even $d$ 
show some differences also at the level of algebraic number theory. 

The remaining group elements are given by $d^2$ displacement operators which we 
write interchangeably as $D_{i,j}$ and $D_{\bf p}$, with the understanding that 
${\bf p}$ is a two-component `vector' with components $i,j$ that 
are integers modulo $d$. The displacement operators obey $D_{\bf p}^\dagger = 
D_{-{\bf p}}$ and 
\begin{equation} D_{\bf p}D_{\bf q} = \tau^{\langle {\bf p}, {\bf q}\rangle }
D_{{\bf p} + {\bf q}} = \omega^{\langle {\bf p}, {\bf q}\rangle}D_{\bf q}D_{\bf p} \ , 
\end{equation}

\noindent where the exponent is given in terms of the components of the `vectors', 
\begin{equation} {\bf p} = \left( \begin{array}{c} i \\ j \end{array} \right) \ , 
\hspace{3mm} {\bf q} = \left( \begin{array}{c} k \\ l \end{array} \right) \hspace{5mm} 
\Rightarrow \hspace{5mm} \langle {\bf p}, {\bf q}\rangle  = kj - li \ . \end{equation}

\noindent Thus $\langle \ , \ \rangle$ is a symplectic form. An explicit matrix 
representation is 
\begin{equation} (D_{i,j})_{r,s} = \tau^{ij + 2js}
\delta_{r,s+i} \ . \end{equation}

\noindent This representation is essentially unique, once $D_{0,j}$ is 
chosen to be diagonal. 

Frequently we will have displacement operators for dimensions $d$ and $d(d-2)$ occurring 
in the same formula. When necessary to avoid confusion 
operators are supplied with superscripts denoting the dimension in which they act, eg.  
$D^{(d)}_{\bf p}$, $D^{(d-2)}_{\bf p}$, $D^{(N)}_{\bf p}$. In this appendix no superscripts 
are necessary because the dimension is always an arbitrary integer $d$. Occasionally we use 
subscripts for the same purpose, thus $\omega_d$ is the $d$th root of unity whenever this is not 
obvious. 

If $F$ is a $GL(2, {\mathbb Z}_d)$ matrix, that is 
to say a $2\times 2$ matrix with entries that are integers modulo $d$, then we 
find when we calculate in modulo $d$ arithmetic that 
\begin{equation} \langle F{\bf p}, F{\bf q}\rangle =  \langle {\bf p}, {\bf q}
\rangle \det{F} \ . \end{equation}

\noindent The condition $\det{F} = 1$ defines the symplectic subgroup $SL(2,{\mathbb Z}_d)$. 
This group is part of the unitary automorphism group of the Weyl--Heisenberg 
group, also known as the Clifford group. Every matrix $F \in SL(2,{\mathbb Z}_d)$ is 
represented by a unitary matrix $U_F$. By definition a Zauner operator is associated 
to a matrix of order three and trace equal to $-1$. The 
matrices $F_z$ and $F_a$, corresponding respectively to the `universal' Zauner operator 
and to the `unusual' Zauner operator in dimensions of the form $d = 9k+3$, are 
\begin{equation} F_z = \left( \begin{array}{cc} 0 & d-1 \\ 1 &  - 1 \end{array} \right) \ , 
\hspace{8mm} F_a = \left( \begin{array}{cc} 1 & 3 \\ 3k & d-2 \end{array} \right) \ . 
\label{Zaunermat} \end{equation}

\noindent See refs.~\cite{Scott, ACFW} for more. Matrices with $\det{F} = - 1$ are 
represented as anti-unitary operators, 
and as such belong to the extended Clifford group~\cite{Marcus}. 

\vspace{1cm}

\section{Parity operators}\label{sec:parity}

\vspace{5mm}

\noindent The symplectic group contains a special involution of order 2, whose 
unitary representative is known as the parity operator, 
\begin{equation} F = 
{\small \left( \begin{array}{rr} -1 & 0 \\ 0 & -1 \end{array} \right) }
\hspace{5mm} \Rightarrow \hspace{5mm} U_F \equiv P \ . \end{equation}

\noindent If $d$ is odd this is a unitary Hermitian operator with spectrum 
$((d+1)/2, (d-1)/2)$. When $d$ is odd the integer $2$ has a multiplicative 
inverse $2^{-1}$ in arithmetic modulo $d$, and we can calculate that  
\begin{equation} \mbox{Tr}D_{\bf p}P = \mbox{Tr}PD_{2^{-1}{\bf p}}P^2D_{2^{-1}{\bf p}}P = 
\mbox{Tr}D_{2^{-1}{\bf p}}PD_{-2^{-1}{\bf p}} = \mbox{Tr}P = 1 \ . \end{equation}

\noindent Hence the parity operator can be expanded as 
\begin{equation} P = \frac{1}{d}\sum_{\bf p}D_{\bf -p} \ . \label{parity} \end{equation}

\noindent Conjugating with the Weyl-Heisenberg group we obtain $d^2$ parity operators 
belonging to the Clifford group. They are the displaced parity operators used in 
Section~\ref{sec:MUB}, and were called phase point operators by Wootters~\cite{Wootters}. 

It is a property of SIC overlap phases that the generalized parity operator 
$P_{\theta}$ occurring in Eq. (\ref{genparity}) is isospectral with the parity 
operator $P$~\cite{Dardo}, but $P_{\theta}$ does not belong to the Clifford group. 

\vspace{1cm}

\section{The Chinese remainder theorem}\label{sec:CRT}

\vspace{5mm}

\noindent We are interested in dimensions of the form $N = d(d-2)$. When $N$ is odd 
$d$ and $d-2$ are relatively prime integers. A theorem from elementary number theory 
then comes into play: the Chinese remainder theorem states that if $n_1$ and $n_2$ are 
relatively prime then any integer $r$ modulo $N = n_1n_2$ can be uniquely expressed 
in terms of a pair of integers $r_i = r$ mod $n_i$ as 
\begin{equation} r = r_1 n_2n_2^{-1} + r_2n_1n_1^{-1} 
\ . \end{equation}  

\noindent Throughout, $n_2^{-1}$ ($n_1^{-1}$) denotes the inverse of the integer 
$n_2$ ($n_1$) in arithmetic modulo $n_1$ ($n_2)$. The formula expresses a ring isomorphism 
between ${\mathbb Z}_N$, the ring of integers modulo $N$, and the ring ${\mathbb Z}_{n_1}\times 
{\mathbb Z}_{n_2}$. This was appreciated in ancient China because it allows arithmetic 
modulo a large integer $N$ to be carried out modulo the smaller integers $n_1$ and 
$n_2$, and the end result reconverted to an integer modulo $N$. The application to 
Weyl--Heisenberg groups as an approach to the SIC problem was pioneered by David Gross~\cite{monomial}.

The Chinese remainder theorem can be used to express the isomorphism between the 
corresponding cyclic groups, and also the isomorphism $H(N) = H(n_1)\times H(n_2)$. We use 
$\omega = e^{\frac{2\pi i}{N}}$ to represent $H(N)$. There holds 
\begin{equation} \omega = \omega_{n_1}^{n_2^{-1}}\omega_{n_2}^{n_1^{-1}} 
\ . \end{equation}

\noindent Namely
\begin{equation} e^{\frac{2\pi i}{N}} = e^{\frac{2\pi i}{n_1n_2}\cdot 1} = 
e^{\frac{2\pi i}{n_1n_2}(n_2n_2^{-1} +n_1n_1^{-1})} = 
e^{\frac{2\pi i}{n_1}n_2^{-1}}e^{\frac{2\pi i}{n_2}n_1^{-1}} \ . 
\label{compomega} \end{equation}

\noindent Given that $\omega_1$ is a primitive root of unity, so is 
$\omega_1^{n_2^{-1}}$, so it would be possible to use this to represent 
$H(n_1)$. However, we choose not to. We then find that 
\begin{equation} D_{i,j} = D^{(n_1)}_{i_1,n_2^{-1}j_1}\otimes D^{(n_2)}_{i_2,n_1^{-1}j_2} 
\ , \label{CR} \end{equation}

\noindent where the matrix representation is, say, 
\begin{equation} D^{(n_1)}_{i_1,n_2^{-1}j_1} = \omega_1^{(2n_2)^{-1}i_ij_i + n_2^{-1}j_1s_1}
\delta_{r_1,s_1+i_1} \ . \end{equation}

\noindent The subscripts on the indices are superfluous, since the aritmetic used for 
the indices is automatically modulo $n_1$. Using vector notation we write 
\begin{equation} D_{\bf p} = D^{(n_1)}_{H_1{\bf p}}\otimes D^{(n_2)}_{H_2{\bf p}} \ , 
\end{equation} 

\noindent where 
\begin{equation} H_1 = \left( \begin{array}{cc} 1 & 0 \\ 0 & n_2^{-1} \end{array} 
\right) \ , \hspace{8mm} H_2 = \left( \begin{array}{cc} 1 & 0 \\ 0 & n_1^{-1} 
\end{array} \right) \ . \end{equation} 

\noindent The Clifford group also splits into a direct product. One finds 
\begin{equation} U_F = U^{(n_1)}_{F_1}\otimes U^{(n_2)}_{F_2} 
= U^{(n_1)}_{H_1FH_1^{-1}}\otimes U^{(n_2)}_{H_2FH_2^{-1}} \ . 
\label{Cliffordsplit} \end{equation}

\noindent Now we specialize to $n_1 = d$, $n_2 = d-2$. Then  
\begin{equation} n_2^{-1} \ {\rm mod} \ n_1 \ = n_1^{-1} \ {\rm mod} \ n_2 = 
\frac{d-1}{2} 
\equiv \kappa \ , \end{equation}

\noindent where the integer $\kappa$ was defined in the last step. (Proof: Calculating 
modulo $d-2$ we find $d(d-1)/2 = 
2(d-1)/2 = d-1 = 1$. The point is that $(d-1)/2$ is an ordinary integer. 
{\it Mutatis mutandis} when calculating modulo $d$.) Thus 
\begin{equation} H \equiv H_1 = H_2 = \left( \begin{array}{cc} 1 & 0 \\ 0 & 
\kappa \end{array} \right) \ . \end{equation}

\noindent For the symplectic matrices one finds 
\begin{equation} F = \left( \begin{array}{cc} \alpha & \beta \\ \gamma & \delta 
\end{array} \right) \hspace{5mm} \Rightarrow \hspace{5mm} 
HFH^{-1} = \left( \begin{array}{cc} \alpha & \kappa^{-1}\beta \\ \kappa \gamma & \delta 
\end{array}\right) \ , \label{Cliffordremainder}
\end{equation}

\noindent where we decide on the modulus in the last step. 

In conclusion, in dimensions $N = d(d-2)$ with $d$ odd the Weyl--Heisenberg group 
allows us to express the Hilbert space as ${\mathbb C}^N = {\mathbb C}^d\otimes {\mathbb C}^{d-2}$ 
in a preferred way.

\end{document}